# *Can 'Robots Won't Save Japan' Save Robotics? Reviewing an Ethnography of Eldercare Automation*

Andrew Hundt, Carnegie Mellon University <ahundt@cmu.edu>

## Abstract

Imagine activating new robots meant to aid staff in an elder care facility, only to discover the robots are counterproductive. They undermine the most meaningful moments of the jobs and increase staff workloads, because robots demand care too. Eventually, they're returned. This vignette captures key elements of James Adrian Wright's ethnography, *"Robots Won't Save Japan"*, an essential resource for understanding the state of elder care robotics. Wright's rich ethnographic interviews and observations challenge the prevailing funding, research, and development paradigms for robotics. Elder care residents tend to be Disabled, so this review article augments Wrights' insights with overlooked perspectives from Disability and Robotics research. This article highlights how care recipients' portrayal suggests that Paro, a plush robot seal, might perform better than the care team and author indicated— leading to insights that support urgent paradigm shifts in elder care, ethnographic studies, and robotics. It presents some of the stronger technical status quo counter-arguments to the book's core narratives, then confronts their own assumptions. Furthermore, it explores exceptional cases where Japanese and international roboticists attend to care workers and recipients, justifying key arguments in Wright's compelling book. Finally, it addresses how *"Robots won't save Japan"* will save Robotics.

## Introduction

*Robots Won't Save Japan* is James Adrian Wright's sincere invitation to Roboticists, Researchers, Corporate Managers, and Policymakers to substantively engage with the messy reality and internal contradictions of modern Robotics Research and Development (R&D) (Wright, 2023). Wright tackles one of the biggest motivating claims for robotics research: that an 'aging society' necessitates the urgent development of care robots. Today's global population as a whole remains demographically stable across age groups as a percentage of the population (Population Pyramid, 2023; UN, 2022). However, countries like Japan anticipate a surge in the proportion of elderly people who require assistance with daily activities like eating, cleaning, and hygiene. Proponents of a robotic model of care often emphasize estimates of the strain elder care could place on budgets and national productivity. They propose robots as a solution, and have even gone so far as to declare that "*Robots Will Save Japan*" (Nakayama, 2006; Kishi, 2011). Roboticists deploy this 'aging society' argument so pervasively that a recent study by Brandão (2021) found it to be present in 30% of all papers across a range of top robotics publications. By contrast, Wright makes a formidable argument that "despite [the] considerable hype, lofty expectations, and substantial investment, robots alone cannot yet deliver on the promise of solving care crises in Japan or elsewhere" (Wright, 2023).

To substantiate his argument, Wright conducted 18 months of ethnographic fieldwork in Japan between 2016 and 2020, embedding with and interviewing a range of key stakeholders. This includes workers and managers at a mid-size care facility he calls Sakura. He also worked with researchers, developers, and managers who work on care robots. One strength of the book is its detailed examination of care workers' knowledge and expertise. Another is its elegant juxtaposition of very different stakeholders' in-depth perspectives and priorities. He analyzes his findings in conjunction with the academic literature and Japan's national political discourse in a manner that considers the unique sociotechnical circumstances of Japan. Wright's analysis proves timely, relevant, and broadly applicable across national, corporate, and academic boundaries, as I will elaborate below.

## The Reality of Elder Care Robotics Research and Development

Wright observes that political and robotics research narratives about the purported urgency of demographic challenges have proven to be powerful tools for the promotion of a "technology-centered reimagining of care, [in which] machines with names like Hug, HAL, and Pepper are intended to supplement or replace the work of human caregivers while also assisting older people to look after themselves, breaking down the tasks of care into linear strings of simple, repeatable physical actions and speech that can be digitally reproduced by computational algorithms and performed by robots" (Wright, 2023). Wright reveals a plethora of internal contradictions within robotics research in practice, in spite of the narrative urgency of elder care robots. Notably, he finds that Roboticists rarely engage directly with care staff or residents deeply enough to meaningfully and economically address physical tasks in practice.

Wright finds that none of the widely-deployed robots he examines address the core objectives set forth in Japan's national Research and Development (R&D) funding goals. In fact, the work of handling and caring for each of these robots increases the labor of front-line care workers at Sakura while undermining job satisfaction and personal connections that come in part from the intimacy of direct tactile care of residents. The robots are also simply too expensive or not designed for realistic operational environments.

For example, the person-lifting robot named Hug necessitates a robot transfer (relocation) as a prerequisite to every transfer of a person. The robot itself can become an obstruction when its design proves too large for existing rooms and hallways. Robot-made traffic jams can bring the buzz of energetic staff and resident activity to a screeching halt. Furthermore, the robot neither accounts for the physical diversity of human bodies (Wright, 2023; Hendren 2020), nor the way robots reduce the opportunities for the social joking, helping, and gossip that typically surround transfer (Wright, 2023), which is one of the essential ways care staff can positively impact residents' and their own quality of life. Wright describes how "Hug threatened the ethical basis of care [and] seemed not only to disrespect older residents but indeed for some care workers, in an almost symmetrical manner, to diminish the value and dignity of their own work and the pride they took in it" (Wright, 2023). Wright's description of the impact of these robot designs viscerally evokes the image of Liz Jackson's Disability Dongle: "Contemporary fairy tales that appeal to the abled imagination by presenting a heroic designer-protagonist whose prototype provides a techno-utopian (re)solution to the design problem. Disability Dongle rhetoric instills […] the value of a quick fix over structural change, thus preventing [people] from seeking out, participating in, and contributing to existing inquiry" (Jackson, Haagaard, & Williams, 2022).

## Disability and Elder Care Robotics

While Wright's analysis is evocative and compelling, the book could have significantly enriched its discourse by integrating insights from Disability literature (Hayashi & Okuhira, 2001; Jackson, Haagaard, & Williams, 2022; Shew, 2023; Dolmage 2017; Shaw et al., 2023; Hendren 2020; Charlton, 1998) and adhering to the principle of "Nothing About Us Without Us" (Charlton, 1998; Hundt et al., 2022). Elder care facilities inherently cater predominantly, if not entirely, to Disabled people. Nonetheless, the book falls short with respect to examining residents' perspectives from their own standpoint, and from the standpoint of the wider Disability community; instead, it skims the surface, focusing on residents' superficial behaviors. Conducting analysis through the lens of Disability presents an extraordinary opportunity to enhance the explanatory power and draw more robust conclusions from this kind of work.

The book describes Ito, a resident with Dementia who is fond of animals and formed a strong bond with Paro, a plush baby seal robot designed for therapeutic use. This story is emblematic of the broader issue where the residents' perspectives might not have been fully considered, leading to potential stigmatization. It emphasized the large amount of time Ito spent with Paro, how she would want to eat with Paro, take it back to her room to be alone, and she would speak to people less often. The care team held a range of perspectives on Ito's interactions with Paro. Some viewed Paro as "really good for Ito" and 'acting as a form of "stimulation"' (Wright, 2023, p. 112). Others worried she might not socialize with people adequately, even expressing "distaste and unease" about her interactions. People in the latter camp "worried that [Paro] was making Ito's Dementia worse by encouraging her to retreat physically into her own room, and mentally into her own world" (Wright, 2023, p. 111).

The application of analyses to individuals has the potential to have enormous impacts, positive and negative, on both a broad spectrum of residents, and on the perceived feasibility of various robot models. Cases of stigmatization and deficit framing of peoples' experiences or behaviors are known to have harmful effects (Bottema-Beutel et al., 2021; Natri et al., 2023; Shaw et al., 2023). Such approaches can strip residents of their agency, particularly when alternative explanations for their experiences are never considered, let alone ruled out. Consider how other care home residents have found Paro to be calming, to foster a sense of companionship, and to aid with social connections (Hung et al., 2021). Perhaps this is how Ito felt. Maybe she was also experiencing social stigma or unmet sensory needs (Houston & Houston, 2015). Such possibilities were not discussed.

Ideally, a well-informed care team could be equipped to respect residents' agency and preferences while supporting the full diversity of their human needs. This includes anticipating the potential for unrecognized sensory needs (Doherty, 2023; Houston & Houston, 2015) as well as unrecognized cognitive or physical disabilities. For instance, it is well-known that many undiagnosed Autistic residents live in elder care facilities like Sakura (Pukki et al., 2023), and there are overlaps in sensory experiences with Dementia (Örulv, 2023; Houston and Houston, 2015). Autistic people are also more likely to develop dementia early (Vivanti 2021), with 30% reporting cognitive decline by middle or older ages on dementia screeners according to Klein et al. (2023). I will briefly explore some overlapping aspects of Autism and Dementia as a specific illustrative example of the general principle. However, individual needs and experiences vary, so what follows might not apply to any specific resident in the book.

Residents with Autistic traits are likely to have intense, but appropriate, 'special interests' that exceed the interest levels typical of non-Autistic residents. Many Autistic people and people with Dementia have very different sensory experiences from those with neither (Doherty, 2023; Wiens, 2014). A wide range of often addressable details such as lights or background noise can continuously demand attention, which can quickly become overwhelming and dysregulating (Price, 2023; Houston and Houston, 2015). Furthermore, social interactions can often differ harmlessly, but even then people are often expected to modify their style of interaction so that they are perceived as socially acceptable for the comfort (e.g. freedom from "distaste and unease") of non-disabled people. Such efforts to hide Autistic traits are known as 'masking' or 'camouflaging', and can be an intense drain on their 'social battery', while also presenting mental health risks to Autistic people (Pukki et al., 2023). In medical settings, Autistic people need SPACE instead— an acronym for Sensory needs, Predictability, Acceptance, Communication, and Empathy (Doherty, 2023), and similar guidance is provided by and for people with dementia (Houston & Houston, 2015; Wiens, 2014), although there are important differences that are out of the scope of this article. To this end, activities like 'stimming', also known as self-stimulation, and recharging via more alone time can be helpful approaches to improve self-regulation and wellbeing, perhaps even with Paro for those who enjoy it.

Therefore, in the case of an undiagnosed Autistic resident, an intense interest in Paro might be an indicator that Paro is providing significant and appropriate relief. Paro might also provide people with a helpfully soothing and self-regulating way to stim, and a communication mechanism for residents to engage in self-advocacy, without speaking, by signaling their need for alone time. If the experiences of care home residents with Paro mirror those observed in hospital experiences for individuals with co-occurring dementia, one might expect expressions of joy, friendship, and social connection, along with displays of personal agency (Hung, 2021) for those who enjoy it. Paro might also support such residents' agency by providing a plausible reason that they can use to remove themselves from the overstimulating and overwhelming environments that care teams are currently unlikely to understand or be able to address (Shaw et al., 2023).

Even if they have the best of intentions, real care teams unfortunately tend to be uninformed about the full spectrum of disabilities (Pukki et al., 2023; Shaw et al., 2023). Decisions tend to be made about Disabled people based on disability stigma, discrimination, and situations that are the result of a lack of reasonable accommodations (Pukki et al., 2023; Shew et al., 2023). Consequently, this approach too often leads to harmful outcomes that are out of line with the actual variability of human needs, social needs, and preferences. Viewed through the lens of disability rights, a decision to remove Paro and push socialization after it is introduced might in some cases be accurately described as denying such residents deserved autonomy, eliminating a significant and appropriate source of relief, all while imposing additional dysregulating stressors.

This discussion of sensory and self-advocacy support needs is just one high impact example that suggests better consideration for the full diversity of people who actually exist in elder care facilities. It illustrates the urgent need for care teams, anthropologists, roboticists, and other researchers to more deeply examine the implications of disability literature in general (Shew (2023) and Dolmage (2017) offer excellent introductions), as well as with respect to robots in particular. Such steps will illuminate the necessity of comprehensive changes to human processes, elder care, and robot design. I also suggest that future work examine ideologically loaded deficit descriptions of Disabled residents, address the common ableist external assumptions that get made about the internal experiences of Disabled people (Pukki et al., 2023; Bottema-Beutel et al., 2021; Monk, Whitehouse, & Waddington, 2022; Bottema-Beutel et al., 2023), and adopt inclusive language guidelines (Bottema-Beutel et al., 2021; Natri et al. 2023).

Future investigations that engage with the broader disability literature might elucidate and substantiate a wide range of essential perspectives on the interaction between care recipients, care personnel, and roboticists that are novel to the academic literature. Furthermore, combining such analyses with thoughtful interactions and input from a broad range of residents could serve as a method to learn about their perspectives, goals, and desires, from their point of view. This approach can substantiate, or refute, such analyses. Appropriate disability-aware methods should be considered for future studies about any interaction between older and/or Disabled populations and the professional disciplines relevant to this article.

## The Status Quo, Politics, and Future of Elder Care Robotics

I also anticipate several arguments from proponents of the current Robot R&D status quo, even in light of the technical and cost limitations highlighted by Wright. For example, Decades of research were necessary for computer vision and natural language processing to develop into practical, commercially viable methods. Some argue that economies of scale significantly reduced costs over that period through capitalist processes of 'creative destruction' (Schumpeter, 1942/2013). Furthermore, proponents may argue that humans need a much better understanding of ways to handle the basic physical motions of robots in complex environments as a necessary prerequisite to more comprehensive and systematic robot designs.

Wright does indicate that future robots might be more capable than those considered in his study. He also poignantly observes capabilities that are entirely out of scope for both currently available and currently planned robots. For example, care workers often sense subtle changes in the way residents interact as early warnings of potential health crises (Wright, 2023). Narratives in favor of robotic care might concede that such essential but indirect services are currently out of scope, then suggest they will become feasible as the technology improves. They might argue that robots also will, in turn, eventually support people to live happy and independent lives for longer. New capabilities such as more powerful and smaller motors, batteries, fabrication methods, and 'Artificial Intelligence' algorithms certainly do have significant potential to address some of the concerns raised by Wright. However, when such narratives depend too heavily on unsubstantiated hypotheticals and do not address the embedded assumptions and flaws of current and planned products, as per the cases Wright describes throughout his book, it too often leads to the "democratization" (read: proliferation) of disability dongles, unjust research, and ineffective products (Jackson, Haagaard, & Williams, 2022; Birhane et al., 2022; Wright, 2023). Such harmful cases are examples of what Shew (2023) terms technoableism, which is a world away from the ideal of democratizing robotics and elder care in the sense of governance by the people and for the people.

Serious scientific study of robots should assess robot-free alternatives and each policy's potential positive and negative impacts and externalities across the short term and the long term. For example, the strengths of Paro that we discussed earlier are promising, but that does not imply Paro should necessarily be introduced, or that robots are generally an adequate solution to the problems experienced by elder care recipients. Despite its benefits, barriers to Paro's use include its cost, increased staff workload, societal stigma, and infection concerns (Hung et al., 2019). Strong supporting evidence is needed before robots are deployed.

One possible alternative scenario to consider beyond speculative assertions about a utopian future of ubiquitous care robots is a crisis in which quality of life is damaged and loneliness increases as robots displace human-to-human contact for care recipients (Wright, 2023; Hsu et al., 2020). Wright notes (Wright, 2023, pp 88, 92, 144) that service quality might also decline if a lower cost and minimally adequate, but inferior, robot substitute to human labor becomes dominant. Wright also poignantly observes how many of the success objectives the robotics field describes for itself risk externalities that could lead to broader devastation due to intensive mining, an intensified climate crisis, and "mountains of toxic electronic and plastic waste" (Wright, 2023). In summary, together with the disability section, this discussion shows that robots won't save *anyone* without the work necessary to ensure a strong method to application and impact fit.

The potency of Wright's ethnographic study is most evident whenever he boldly addresses an often-ignored reality. A striking illustration is when he corroborates the field's political use of national demographic statistics and an 'aging society' narrative to drive Robotics research under the guise of a technical imperative. It is disconcerting to contemplate so many roboticists and policymakers evading this enduring truth: the world has a globally stable population that is expected to age gracefully for the foreseeable future (Population Pyramid, 2023; UN, 2022); and immigration (Sakanaka, 2007), unlike robots, is an approach to mitigating demographic concerns that, while politically complex, is technically feasible today. This reality has been conspicuously sidelined in countless documents and discussions. One plausible explanation is the pragmatic political support of nations' geopolitical status quo for the purpose of appropriating funds to Robotics research.

Wright indicates that "on an ideological level, a pro-(im)migration policy appears to reflect a globalist outlook, 'opening up' Japan; a pro-robot policy seems to represent the techno-nationalist and xenophobic ideology of a technologically closed country" (Wright, 2023). Grappling with these uncomfortable possibilities, and problematic motives in general, could serve to improve the research direction, goals, and outcomes of robotics as a field. Similarly, elder care robotics research in Japan has been reported to feature one-dimensional stereotyping, paternalism, and technocentric tendencies (Hsu et al., 2020). Their findings are familiar, as I regularly encounter research authored across numerous countries on the topic of Disability in Robotics that contains harmful perspectives. Wright's comprehensive exploration of elder care robotics in Japan raises critical concerns about the prevalence of motivated reasoning in robotics research worldwide, a topic that warrants deeper investigation.

Even so, no one lens can capture the entirety of a country, industry, or research field. Japanese roboticists are not a monolithic population (Hsu et al., 2020), nor are roboticists in general (Bhattacharjee et al., 2019; Hundt et al., 2022). For example, Hsu et al. (2020) conducted interviews of elder care roboticists in Japan, encountering people who highlighted unintended consequences of robots, expressed critiques highlighting possible negative outcomes such as loneliness, recognized the wasted resources when the perspectives of elderly people are not considered, and described being denied the resources necessary to engage with end users by their organizations even though they wished to do so. Wright's work could have benefited from further engagement with such marginalized perspectives, either directly or via existing literature. At the same time, Wright's critiques have merit with respect to mainstream norms for roboticists (Brandão, 2021), as noted in the introduction. Robotics as a field would be better off if it onboarded thoughtful and high quality marginalized perspectives to the mainstream, particularly considering the substantial increase in research rigor such a shift would entail.

## Conclusion

As a whole, an effective way for roboticists to approach this fantastic book is to do so with humility and an open mind receptive to meaningful critiques. Prioritize the needs of care recipients first, and recall that one can simultaneously disagree with and learn from a range of perspectives, as needed. Many of the issues raised by Robots Won't Save Japan resonate with my own experience of robot care research and literature outside of Japan.

*Robots Won't Save Japan* is essential reading for roboticists who want a better grasp of the problems they are working on. It provides a window through the veil of automation bias that so often clouds our judgment. It elucidates rarely acknowledged conflicts of interest among Disabled populations, care teams, and the robotics R&D teams that purport to serve them.

This article, in turn, has shown how the perspectives of Disabled people make or break the viability and conclusions of elder care research across disciplines. It is essential to center the perspectives of Disabled people, motivating the urgent need for a paradigm shift in elder care, ethnographic research, robotics, and policymaking to center Disabled people and respect the principle of "Nothing About Us Without Us". It demonstrates that robotics communities already exist with a better culture of care, and they need to be amplified. Together with *Robots Won't Save Japan*, there is an enormous opportunity and motivation to adopt new research questions and improve the funding, implementation, and methods necessary for the responsible conduct of Robotics research. While the book might not save robotics on its own, it can certainly help. In conclusion, Robots Won't Save Japan is a must-read for Roboticists, Researchers, Managers, Executives, Policymakers, and anyone else with an interest in robots for care related research.

## Acknowledgements

Thank you to Maki Kaneko for input regarding Japan and Disability. Thanks to Maria Ryabova at the University of Pittsburgh for valuable insight into anthropological research methods. Finally, thank you to everyone who provided guidance on authoring academic book reviews and/or feedback on the article. LLMs were used to facilitate selected writing style and grammar adjustments, as well as to format references.

## References

Birhane, A., Pratyusha Kalluri, Dallas Card, William Agnew, Ravit Dotan, & Michelle Bao. (2022). The Values Encoded in Machine Learning Research. In Proceedings of the 2022 ACM Conference on Fairness, Accountability, and Transparency (FAccT '22). Association for Computing Machinery, New York, NY, USA, 173–184. https://doi.org/10.1145/3531146.3533083

Bottema-Beutel, K., Kapp, S. K., Lester, J. N., Sasson, N. J., & Hand, B. N. (2021). Avoiding Ableist Language: Suggestions for Autism Researchers. Autism in Adulthood. Mar 2021.18-29. http://doi.org/10.1089/aut.2020.0014

Bottema-Beutel, K., Kapp, S. K., Sasson, N., Gernsbacher, M. A., Natri, H., & Botha, M. (2023). Anti-ableism and scientific accuracy in autism research: a false dichotomy. Front. Psychiatry. 14:1244451. doi: https://doi.org/10.3389/fpsyt.2023.1244451

Charlton, J. I. (1998). Nothing about us without us: disability oppression and empowerment. University of California Press, Berkeley.

Doherty M., McCowan S., Shaw S. C. K. (2023). Autistic SPACE: A novel framework for meeting the needs of autistic people in healthcare settings. British Journal of Hospital Medicine, 84(4), 1–9. https://doi.org/10.12968/hmed.2023.0006

Dolmage, J. (2017). Academic Ableism: Disability and Higher Education. University of Michigan Press, Ann Arbor. https://doi.org/10.3998/mpub.9708722

Eric L. Hsu, Anthony Elliott, Yukari Ishii, Atsushi Sawai, & Masataka Katagiri. (2020). The development of aged care robots in Japan as a varied process. Technology in Society, Volume 63, 101366, ISSN 0160-791X, https://doi.org/10.1016/j.techsoc.2020.101366

Hendren, S. (2020). What Can a Body Do?: How We Meet the Built World. Riverhead Books.ISBN 9780735220003

Houston, A. & Houston, D. (2015). Dementia & Sensory Challenges [Leaflet and Video]. Life Changes Trust. Retrieved from https://www.alzscot.org/sites/default/files/2019-07/Sensory%20Leaflet.pdf and https://youtu.be/U_L2107EweQ

Hundt, A., William Agnew, Vicky Zeng, Severin Kacianka, & Matthew Gombolay. (2022). Robots Enact Malignant Stereotypes. (2022) ACM Conference on Fairness, Accountability, and Transparency (FAccT '22), June 21– 24, 2022, Seoul, Republic of Korea. ACM, New York, NY, USA, 14 pages. https://doi.org/10.1145/3531146.3533138

Hung, L., Liu, C., Woldum, E. et al. The benefits of and barriers to using a social robot PARO in care settings: a scoping review. BMC Geriatr 19, 232 (2019). https://doi.org/10.1186/s12877-019-1244-6

Hung L, Gregorio M, Mann J, et al. Exploring the perceptions of people with dementia about the social robot PARO in a hospital setting. Dementia. 2021 ;20(2):485-504. DOI: https://doi.org/10.1177/1471301219894141

Jackson, L., Haagaard, A., & Williams, R. (2022, April 19). Disability Dongle. The CASTAC Blog. https://blog.castac.org/2022/04/disability-dongle/

Kishi, N. (2011). Robotto ga nihon wo sukuu [Robots will save Japan]. Tokyo: Bungeishunjū.

Klein, C. B., McQuaid, G. A., Charlton, R. A., Klinger, L. G., & Wallace, G. L. (2023). Self-reported cognitive decline among middle and older age autistic adults. Autism Research, 16(3), 605–616. https://doi.org/10.1002/aur.2877

M. Brandão. (2021). Normative roboticists: the visions and values of technical robotics papers. In 2021 30th IEEE International Conference on Robot & Human Interactive Communication (RO-MAN), Vancouver, BC, Canada, 671-677. https://doi.org/10.1109/RO-MAN50785.2021.9515504

Monk, R., Whitehouse, A. J. O., Waddington, H. (2022). The use of language in autism research. Trends in Neurosciences, Volume 45, Issue 11, Pages 791-793, ISSN 0166-2236, https://doi.org/10.1016/j.tins.2022.08.009

Nakayama, S. (2006). Robotto ga nihon o suku'u [Robots Will Save Japan]. Tokyo: Tōyō Keizai Shinpōsha.

Nanavati, A., Alves-Oliveira, P., Schrenk, T., Gordon, E. K., Cakmak, M., & Srinivasa, S. S. (2023). Design Principles for Robot-Assisted Feeding in Social Contexts. In Proceedings of the 2023 ACM/IEEE

Natri, H. M., Abubakare, O., Asasumasu, K., Basargekar, A., Beaud, F., Botha, M., Bottema-Beutel, K., Brea, M. R., Brown, L. X. Z., Burr, D. A., Cobbaert, L., Dabbs, C., Denome, D., Des Roches Rosa, S., Doherty, M., Edwards, B., Edwards, C., Liszk, S. E., Elise, F., Fletcher-Watson, S., Flower, R. L., Fuller, S., Gassner, D., Onaiwu, M. G., Good, J., Grant, A., Haddix, V. L., Heraty, S., Hundt, A., Kapp, S. K., Keates, N., Kulshan, T., Lampi, A. J., Latimer, O., Leadbitter, K., Litton Tidd, J., Manalili, M., Martin, M., Millichamp, A., Morton, H., Nair, V., Pavlopoulou, G., Pearson, A., Pellicano, L., Porter, H., Poulsen, R., Robertson, Z. S., Rodriguez, K., Roux, A., Russell, M., Ryan, J., Sasson, N., Smith Grier, H., Somerville, M., Sorensen, C., Stockwell, K. M., Szymanski, T., Thompson-Hodgetts, S., van Driel, M., VanUitert, V., Waldock, K., Walker, N., Watts, C., Williams, Z., Woods, R., Yu, B., Zadow, M., Zimmerman, J., Zisk, A. H. (2023). Anti‑ableist language is fully compatible with high‑quality autism research: Response to Singer et al.(2023). *Autism Research*, 16(4), 673-676. https://doi.org/10.1002/aur.2928
https://psyarxiv.com/3vjnq

International Conference on Human-Robot Interaction (HRI '23) (pp. 24–33). Association for Computing Machinery. https://doi.org/10.1145/3568162.3576988

Population Pyramids of the World and Japan from 1950 to 2100. (2023). Retrieved August 23, 2023, from https://www.populationpyramid.net/world/2023/ and
https://www.populationpyramid.net/japan/2023/

Price, D.V. (2022). Unmasking Autism: The Power of Embracing Our Hidden Neurodiversity. Octopus Publishing Group. ISBN-13 9780593235232

Pukki, H. et al. (2022). Autistic Perspectives on the Future of Clinical Autism Research. Autism in Adulthood. Jun 2022.93-101. http://doi.org/10.1089/aut.2022.0017

Reiko Hayashi & Masako Okuhira. The Disability Rights Movement in Japan: Past, present and future. https://doi.org/10.1080/09687590120083994

Sakanaka, H. (2007). The Future of Japan's Immigration Policy: A Battle Diary. Asia-Pacific Journal: Japan Focus, 5(4), 1–9.

Schumpeter, J. A. (1942, 2013). Capitalism, Socialism, and Democracy. Routledge.

Shaw, S. C., Carravallah, L., Johnson, M., O'Sullivan, J., Chown, N., Neilson, S., & Doherty, M. (2023). "Barriers to healthcare and a 'triple empathy problem' may lead to adverse outcomes for autistic adults: A qualitative study." *Autism*. https://doi.org/10.1177/13623613231205629

Shew, A. (2023). Against Technoableism: Rethinking Who Needs Improvement. W. W. Norton, ISBN 978-1-324-03666-1 https://wwnorton.com/books/9781324036661

Tapomayukh Bhattacharjee, Maria E. Cabrera, Anat Caspi, Maya Cakmak, & Siddhartha S. Srinivasa. (2019). A Community-Centered Design Framework for Robot-Assisted Feeding Systems. In Proceedings of the 21st International ACM SIGACCESS Conference on Computers and Accessibility (ASSETS '19). Association for Computing Machinery, New York, NY, USA, 482–494.
https://doi.org/10.1145/3308561.3353803

United Nations, DESA, Population Division. (2022). World Population Prospects 2022. Retrieved August 23, 2023, from http://population.un.org/wpp/ ,
https://population.un.org/wpp/Graphs/DemographicProfiles/Pyramid/900

Vivanti, G., Tao, S., Lyall, K., Robins, D. L., & Shea, L. L. (2021). The prevalence and incidence of early-onset dementia among adults with autism spectrum disorder. Autism Research, 14(10), 2189–2199. https://doi.org/10.1002/aur.2590

Wiens, A. (2014, May 13). Understanding behavioral problems in dementia [Video file]. The Royal. Retrieved from https://youtu.be/08pcvdfIJig

Örulv, L. (2023). Neurodiversity and dementia: Pitfalls, possibilities and some personal notes. In Ward, R., & Sandberg, L. (Eds.). (2023). Critical Dementia Studies: An Introduction (312 pages, 19 B/W illustrations). Taylor & Francis. ISBN 9781032118802 https://doi.org/10.4324/9781003221982

Wright, James Adrian (2023). *Robots Won't Save Japan: An Ethnography of Eldercare Automation* (The Culture and Politics of Health Care Work) (p. 3). ILR Press, an Imprint of Cornell University Press. Pp 3, 6, 88, 92, 117, 144-145 Kindle Edition. 9781501768040
https://www.cornellpress.cornell.edu/book/9781501768040/robots-wont-save-japan/